\documentclass[twocolumn,prl,aps,amsmath,amssymb,superscriptaddress,showpacs]{revtex4-1}

\usepackage{epsfig}
\usepackage{float}

\def\ai         {{\it ab initio}}
\def\gd         {\delta} 
\def\gee        {\varepsilon}
\def\gl         {\lambda}
\def\go         {\omega}
\def\gql        {{\bf q} \gl}
\def\rr         {{\bf r}}

\def\gS         {\Sigma}
\def\la         {\langle}
\def\ra         {\rangle}

\def\tpa        {{\it trans}--polyacetylene}

\def\zpm        {zero--point motion}
\def\zpme       {zero--point motion effect}
\def\epc        {electron--phonon coupling}
\def\epi        {electron--phonon interaction}
\def\ep         {electron--phonon}
\def\dw         {Debye--Waller}

\def\se         {self--energy}

\def\Sfs        {spectral functions}

\renewcommand{\]}{\right]}
\renewcommand{\(}{\left(}
\renewcommand{\)}{\right)}

\begin{document}
 
\title{Effect of the quantistic zero--point atomic motion on the opto--electronic properties
of diamond and trans--polyacetylene}

\author{Elena Cannuccia}
\affiliation{Dipartimento di Fisica, Universit\`a di Roma ``Tor Vergata''}
\affiliation{European Theoretical Spectroscopy Facility}

\author{Andrea Marini}
\affiliation{Dipartimento di Fisica, Universit\`a di Roma ``Tor Vergata''}
\affiliation{European Theoretical Spectroscopy Facility}
\affiliation{Nano-Bio Spectroscopy group, Dpto. F\'isica de Materiales, Universidad del Pa\'is Vasco, Spain}
\affiliation{IKERBASQUE, Basque Foundation for Science, E-48011 Bilbao, Spain}

\date{\today}
\begin{abstract}
The quantistic zero--point motion of the carbon atoms is shown to induce strong effects on the opto--electronic properties of diamond and \tpa, 
a conjugated polymer.  By using an
\ai\,\,approach, we interpret the sub--gap states experimentally observed in diamond in terms of entangled electron--phonon states.  These states
also appear in \tpa\, causing the formation of strong structures in the band--structure that even call into question the
accuracy of the band theory.  This imposes a critical revision of the results obtained for carbon--based nano--structures by
assuming the atoms frozen in their equilibrium positions.
\end{abstract} 
\pacs{71.38.-k, 63.20.dk, 79.60.Fr, 78.20.-e}
%
%
%
%
\maketitle

Carbon--based nano--structures represent the natural candidates to replace silicon--based 
materials in the devise of efficient opto--electronic devices. Conjugated polymers, in particular,
have been shown to have peculiar properties related to the very fast 
relaxation of charge carriers due to the electron--phonon coupling~\cite{Kersting1993}. This
ultra--fast dynamics has made possible to use conjugated polymers to devise efficient light emitting 
diodes~\cite{J.H.Burroughes1990} or nano--scale optical switches~\cite{IgnacioB.Martini2007}.

Despite the rapid development of technological applications the role
of atomic vibrations in carbon--based nano--structures has been only treated in a semi--empirical manner, 
boosted by the essential question of the mobility of charged carriers in organic devices~\cite{Hannewald2004}.
However, these approaches are based on Hamiltonians that rely on parameters which are difficult to extract 
from experiments and clear--cut conclusions are still elusive. 
In contrast, the most accurate, parameter--free and up--to--date description of the electronic properties of 
bulk and nano--sized materials, is based on \ai\,\,methods. 
These techniques benefit of the predictivity and accuracy of Density--Functional Theory (DFT)~\cite{Onida2002} 
merged with Many--Body Perturbation Theory (MBPT)~\cite{mahan}. 
The goal of the \ai\,\,methods is to describe and predict in a quantitative manner, the opto--electronic
properties of any electronic system, starting from its atomic configuration. 
The result is a wealth of techniques like the GW method~\cite{Onida2002} that has been successfully 
applied to a large number of different systems, among which carbon--based nano--structures.

In the GW approach, as well as in other applications of the \ai\,\,methods, a standard approximation is to assume the atoms 
frozen in their equilibrium positions. 
Many years ago~\cite{Cardona2006}, however, the pioneering works of Heine, Allen and Cardona\,(HAC) pointed 
to the fact that, even when the temperature vanishes, the quantistic zero--point motion of the atoms\,(the \zpme)
can induce large corrections to the 
electronic levels, making purely electronic theories (like the GW method) inadequate.
Nevertheless, the enormous numerical difficulties connected with the calculation of the \epi\,\,has 
{\it de-facto} prevented the systematic application of the HAC theory.
Nowadays, the advent of more refined numerical techniques, has made possible to ground the HAC approach
in a fully \ai\,\,framework~\cite{Marini2008,*Capaz2005}.
More recently, on this journal, Giustino et al.~\cite{Giustino2010,Zollner1992} found a large zero--point renormalization 
(615\,meV) of the band--gap of bulk Diamond.
The HAC approach is, however, based on a static theory of the \epc\,
and, Giustino et al.~\cite{Giustino2010}
in their work, rise a worryingly doubt:
{\em We note that the good agreement between our calculations and experiment may be somewhat fortuitous 
since the HAC theory does not take into account dynamical effects}.

In this work we show, indeed, that when the temperature vanishes, the quantistic zero--point motion of the atoms
induces strong {\em dynamical} effects on the 
opto--electronic properties of Diamond and \tpa, a paradigmatic nano--structure. 
The sub--gap peaks appearing in the experimental absorption spectrum of Diamond are interpreted 
in terms of polaronic states, composed of entangled electron--phonon pairs, that cannot be described by the HAC theory.
When the reduced dimensionality of the system enhances the amplitude of the atomic vibrations,
the \zpme\, even fragments the electrons in a continuum of polaronic states.
This is the case of \tpa, where the single--particle band--structure is replaced by a jelly-like 
electronic distribution and the wave--functions of electrons and atoms are stretched along the polymer axis.
This is the break--down of the band--theory.
By disclosing the physical motivations of the \zpme, we discuss how the present results lead
to potentially ground--breaking consequences on our understanding of the opto--electronic properties 
of carbon--based nano--structures.

The \zpme\,\,can be understood by using simple arguments. At finite temperature the atoms oscillate
around their equilibrium positions. These oscillations can be mapped in a system of non--interacting 
harmonic oscillators. As a consequence of the quantistic nature of the atoms, 
when the temperature goes to zero the atoms collapse in a ground state with a finite energy.
This is the zero--point energy that induces quantum fluctuations of the atoms. The effect of this purely quantistic motion on the electronic levels can be
described using MBPT applied to the \ep\,\,problem\,\cite{mahan}.
It can be shown that, at zero temperature, the electron--phonon \se, taken to the lowest non vanishing 
order of perturbation theory, is composed of two contributions. The first term is the Fan \se
\begin{align}
    \Sigma^{Fan}_{I}(\go)=\sum_{J\gl}\sum_{\xi=\pm} 
\frac {N^J_{\xi}{\left| \la I \right| H_{el-ph} \left| J\ra\otimes\right. \left| \gl\ra \right|}^2} 
{N_q\(\go-\gee_{J} +\xi \go_{\gl} -i0^{+}\)},
\label{eq:se}
\end{align}
with $\left|J\right.\ra$ and $\left|\gl\right.\ra$ the generic electronic and phononic state.
$N^J_{+}=f_J$ and $N^J_{-}=1-f_J$, with $f_J$ the electronic occupations. $N_q$ is the total
number of transferred momenta used to integrate the \se.
The key quantities in Eq.\ref{eq:se} are the \ep\,\,matrix elements 
$g^{\gl}_{IJ}\equiv \la I \left| H_{el-ph} \right.\left| J\ra\otimes \right| \gl\ra$, with
$H_{el-ph}$ the \ep\, Hamiltonian. The $g^{\gl}_{IJ}$
are calculated
\ai\,\,using Density--Functional--Perturbation--Theory\,\cite{baroni2001}. 
A frequency independent \dw\,\,term, $\Sigma^{DW}_{I}$ is added to $\Sigma^{Fan}_{I}$ in order to 
preserve the translational invariance of the theory. More details about the \dw\,\,term can be found, 
for example, in Ref.~\cite{Zollner1992}.

The full frequency--dependent Green's function $G_{I}\(\go\)$ is readily defined to be
$G_{I}(\go)=\(\go-\gee_{I}-\Sigma^{Fan}_{I}(\go)-\Sigma^{DW}_{I}\)^{-1}$.
The true single--particle excitations of the system are obtained as poles of $G_{I}$. The HAC theory and the more
general Quasi--Particle approximation\,(QPA) can be obtained making approximations on the frequency dependence of
$\Sigma^{Fan}_{I}(\go)$.
However we would like to follow a different path.

Physically Eq.\ref{eq:se} describes the scattering of the bare electronic state 
$\left| I\right.\ra$ with the continuum of phonons that surrounds the 
state $\left| J\right.\ra$. These scatterings are weighted by the coupling terms
$g^{\gl}_{IJ}$.
This scenario can be described using the well--known Fano theory\,\cite{fano} which describes, in 
general, the coupling of a discrete state ($\left| I\right.\ra$) with a set of final states
surrounded by a continuum of excitations ($\left| J\right. \ra\otimes \left| \gl\right.\ra$).
The scattering with many possible final states induces interference effects. When
these effects are small the term with $J=I$ dominates the sum in Eq.\ref{eq:se},
and the Fano theory predicts the spectral function\,(SF) $A_{I}\(\go\)\equiv\frac{1}{\pi}\left|\Im\[G_{I}\(\go\)\]\right|$
to be a Lorentzian. This is the QPA, and the center of the Lorentzian is the Quasi--Particle\,(QP) energy, defined
to be 
$\gee_{I}+Z_{I} \gS^{Fan}_{I}(\gee_{I})+\Sigma^{DW}_{I}$, with
$Z_{I}= \( 1 -\left.\frac{\partial \Re\gS^{Fan}_{I}(\go)}{\partial \go}\right|_{\go=\gee_{I}} \)^{-1}$
the renormalization factor~\cite{mahan}.
If in addition $\mid g^{\gl}_{IJ}\mid^2\ll\go_D$, with $\go_D$ the Debye frequency, the Lorentzian width goes to zero,
$Z=1$ and we have the HAC approach. This corresponds to a frequency independent $\gS^{Fan}$.
When the coupling factors $g^{\gl}_{IJ}$ are large the interference effects force the system
to create coherent packets of electron--phonon pairs,
\begin{align}
\left| P\right. \ra=\Lambda^P_I \left| I\right. \ra+\sum_{J\gl}\Xi^P_{J\gl} \left| J \ra \otimes \right| \gl \ra.
\label{eq:polarons}
\end{align} 
We define $\left| P\right. \ra$ a polaronic state 
with energy $E_P$.
Indeed, it can be demonstrated\,\cite{cannuccia_u} that, by choosing the coefficients $\Lambda^P_I$ and $\Xi^P_{J\gl}$ 
as eigenvectors of an effective Hamiltonian, the Green's function corresponding to Eq.\ref{eq:se}, can be rewritten as 
$G_{I}(\go)= \sum_{P} {\left| \Lambda_{I}^{P}\right|}^2 (\go-E_{P}+i0^{+})^{-1}$.
It also follows that $\la P \left|\right. P\ra=1$.
Physically Eq.\,\ref{eq:polarons} describes a system where the electrons are replaced by a continuum of polarons whose
purely electronic part is weighted by $\left| \Lambda_{I}^{P}\right|^2$.
The QPA is recovered when the $\left| \Lambda_{I}^{P}\right|^2$ factors have a Lorentzian distribution around the QP energy. However,
in general, if the $\left| \Lambda_{I}^{P}\right|^2$ are small the second term in the r.h.s. of Eq.\,\ref{eq:polarons}  is large
making the $A_{I}(\go)$ to deviate from the simple Lorentzian lineshape. In this case the electron is fragmented in several 
polaronic states and the QPA is expected to fail.

\begin{figure}[H]
\begin{center}
\epsfig{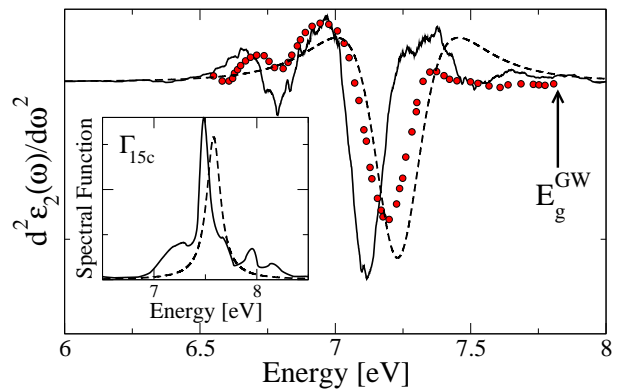}
\end{center}
\caption{\footnotesize{
The experimental\,\cite{Logothetidis1992} $d^2\gee_2\(\go\)/d\go^2$ of bulk Diamond near the 
absorption edge (red spheres) is compared with the QPA (dashed line) and with the full--dynamical theory 
(full line). 
The optical gap is defined by transitions between the $\Gamma'_{25v}$ (occupied) and $\Gamma_{15c}$ (empty) states.
The $\Gamma'_{25v}$ state is a genuine QP state. The $\Gamma_{15c}$ SF, instead (shown in the inset), 
is characterized by strong side--peaks that well reproduce the sub--gap peaks appearing in the experimental spectrum
below $7.1$\,eV. $E^{GW}_g$ represents the QP direct gap obtained by neglecting the \zpme.
}}
\label{fig:diamond}
\end{figure}

In the inset of Fig.\ref{fig:diamond} the SF $A_{I}\(\go\)$ of bulk Diamond,
obtained by using the full dynamical dependence of the \se, is compared with the QPA for the
$\Gamma_{15c}$ state, the bottom of the conduction bands\cite{calculations}. 
The SF, instead of being a simple Lorentzian, shows peaks at $7.25$ and $7.95\,eV$.
These peaks are due to polaronic states $\left| P\right. \ra$ each carrying a fraction 
(given by $\left| \Lambda_{I}^{P}\right|^2$) of the total electronic charge.
These peaks allow to explain the sub--gap states observed experimentally. 
Indeed the $\Gamma_{15c}$ state contributes to the onset of the absorption via transitions with the top of the 
valence bands, represented by the $\Gamma'_{25v}$ state.
The corresponding contribution to the absorption edge will be given by the
convolution of the $\Gamma'_{25v}$ and $\Gamma_{15c}$ SFs. This convolution is used,
in Fig.\,\ref{fig:diamond}, to calculate the second derivative of the dielectric function
and to compare it with the experimental spectrum~\cite{Logothetidis1992}.
The experimental absorption clearly shows the absorption gap at $7.19\,eV$ together
with some sub--gap structures at $6.71$ and $6.94\,eV$.
Such structures are completely absent in 
the QPA or using the HAC approach.
The dynamical theory, instead, leads to an excellent agreement with the experimental result.

\begin{figure} [H]
\begin{center}
\epsfig{figure=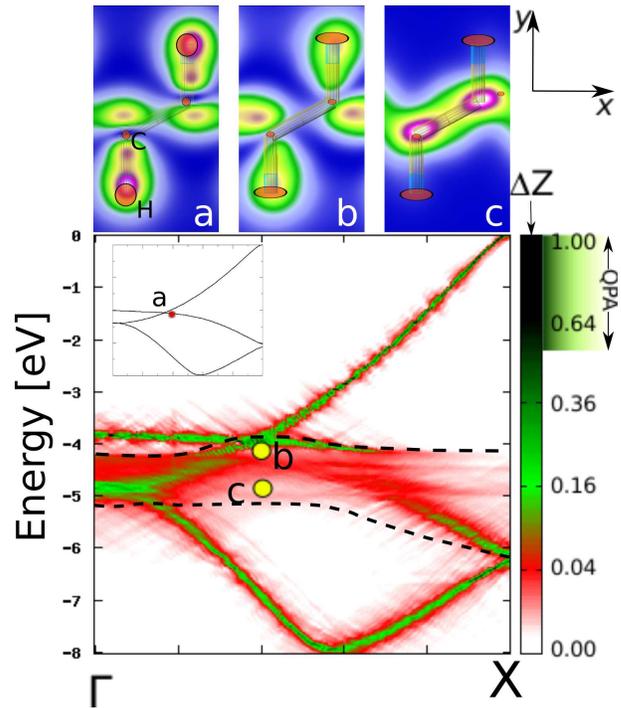,width=8cm}
\end{center}
\caption{\footnotesize{Two--dimensional plot of the SF $A\(k,\go\)$ in the energy region of the last 
three occupied bands. The range of values of $A$ are given in terms of the dimensionless quantity 
$\Delta Z$ (see text) that measures the elemental charge carried by the SF.
Except the $\pi/\pi^*$ states, perpendicular to the polymer plane, the electronic levels acquire a large 
energy indetermination that is particularly dramatic in the region enclosed by the two dashed line. 
This destructive effect is, instead, a cooperative interaction between atoms and electrons that is 
reflected in the stretching of the polaronic states in the direction of the polymer axis.
This is evident from the comparison of the wavefunctions of a reference bare state (left frame, point $a$) with 
two polaronic states (central and right frame, points $b$ and $c$). 
The standard deviation of the atoms in their unperturbed state and in the polaronic states is represented by the 
size of the spheres in the three upper frames.}}
\label{fig:2D}
\end{figure}

This result, while supporting the correctness of a dynamical theory of the \zpme, questions the validity of the QPA 
and of the static HAC approach.
The strength of the dynamical \zpme\,\,in Diamond is unexpected,
and opens the path to potentially stronger effects in carbon--based nano--structures.
Indeed, the strength of the electron--phonon \se\,\,is linked to the amplitude of the atomic vibrations.
In a nano--structure the atoms are more free to oscillate, thanks to the reduced symmetry of the system. 
This can be easily verified by considering a {\em conjugated} polymer, as \tpa.
This is a 1D chain made up of repeated structural units linked by alternated single and double bonds between
the carbon atoms~\cite{HeegerReview}. The unit cell contains 2 carbon and 2 hydrogen atoms, which lay on the 
same plane, as shown in the upper frames of Fig.\ref{fig:2D}.
We can associate an average quantistic size to the atoms by using the 
standard deviation $\sigma_i$ of the atom $i$ in its ground--state wavefunction~\cite{Cardona2006}.
This is $\vec{\sigma}_i\approx\sqrt{\sum_{\gl} \go_{\gl}^{-1} \left |\vec{\xi}_i\(\gl\) \right|^2}$, 
with $\vec{\xi}_i\(\gl\)$ the polarization vector of the phonon mode $\gl$.
In diamond $\sigma_C\approx 0.1\,a.u.$, independently on the direction.
In \tpa\,\,the smaller distance between carbon atoms, slightly reduces the standard deviation along the $\hat x$ direction
where $\sigma_C\approx 0.08\,a.u.$ while in the $\hat y$ direction $\sigma_C\approx 0.16\,a.u.$. 
Hydrogen is twelve times lighter than carbon. As a consequence its standard deviation in the polymer plane is $\sigma_H\approx 0.2\,a.u.$.

To clearly visualize the dramatic effect of the \zpme\,\,on the electronic structure of \tpa\,\,we 
define a global SF $A\(k,\go\)\equiv\sum_{n} \frac{1}{\pi}\left|\Im\[G_{nk}\(\go\)\]\right|$,
where we have explicitly expressed the index $I$ in terms of the band ($n$) and
k--point ($k$) indexes. $k$ is taken in the $\Gamma-X$ direction\,\cite{calculations}. Physically $\Delta Z\equiv A\(k,\go\)\Delta\go$ 
gives the fraction of electronic charge carried by the state with k--point $k$ in the small energy range $\Delta\go$. 
From the definition of $A\( k, \go\)$ it follows that $\Delta Z\propto\left| \Lambda_{I}^{P}\right|^2$.
The $A$ function is a very peaked function of $k$ and $\go$ both in the HAC approach and in the QPA
with the charge confined in very sharp single--particle states. The plot of the $A$ function, shown
in Fig.\ref{fig:2D} in the energy range of the last three occupied bands, gives instead a completely 
different picture.
The intensity reported in the color scale of the Fig.\ref{fig:2D} refers 
to the dimensionless quantity $\Delta Z$ calculated with $\Delta\go=50\,meV$. As a reference
we show the range of values of $\Delta Z$ corresponding to the QPA.
Note that the static HAC theory corresponds to $\Delta Z=1$.

With the exception of the last occupied band, the charge carried by the SF is lower then $0.4$, well 
below the range of values where the QPA or the HAC theory is accurate.
More importantly very small values of $\Delta Z$ mean small $\left| \Lambda_{I}^{P}\right|^2$ and, consequently,
large contributions of the electron--phonon pairs in the polaronic state (second term in the r.h.s of Eq.\ref{eq:polarons}).
This effect is even more dramatic 
in the central energy region, where the bands {\em disappear} being replaced
by an almost uniform, jelly--like charge distribution that, in Fig.\ref{fig:2D}, is represented by the region enclosed
by the two dashed lines.
Physically the system is forced to split the electronic levels in several polaronic states. Single--particle states are not allowed anymore.
This is an unambiguous signature of the breakdown of the band theory.

With respect to more deeper states the last occupied bands, that has a $\pi/\pi^*$ character, 
corresponds to states distributed perpendicular to the polymer axis. As a consequence they
feel less the effect of the on--plane vibrations and almost all electronic charge is confined in sharp quasi--particle--like peaks.

A more careful analysis reveals that the destructive effect of the \zpme\, on the electronic band--structure is, instead,
a signature of a cooperative electron--atom dynamics.
Indeed, Eq.\ref{eq:polarons} allows to calculate, \ai, several properties of the polaronic states. 
By projecting $\left| P\right. \ra$ on the real--space we can define the polaronic wave--function, 
$\la \rr\left| P\right.\ra$, and also calculate the standard deviation 
$\vec{\sigma}_i^P$ of the atom $i$ that participates in the polaronic state.
This can be used to explain the physical properties of the wide region
enclosed by the two dashed lines in Fig.\,\ref{fig:2D}. In the upper frame $(a)$ of Fig.\,\ref{fig:2D} we consider
a bare electronic state that is localized on the $C-H$ bond. The \zpme\,\,splits the charge of this state in several
polaronic states. Two of the most intense are the states $(b)$ and $(c)$ whose wave--functions are shown in
the upper frames of Fig.\,\ref{fig:2D}. By moving towards the center of the region 
the electronic charge is gradually moved from the $C-H$ bond towards the polymer axis. In the state
$(c)$ this charge transfer is more pronounced and the wave--function is completely delocalized along the polymer axis.
Even more stunning is the effect of the \zpm\,\,on the atomic standard deviation. In both
states $(b)$ and $(c)$ $\vec{\sigma}_i^P$ is decreased by $50\%$ in the $\hat y$ direction 
and increased by $150\%$ in the direction of the axis. 
The present results show that the \zpme\, works to
enhance the delocalization of the charge carriers
by stretching both electrons {\em and}
atoms along the polymer axis.
This cooperative effect
agrees with the general interpretation of the high electronic mobility in polymers as dictated by
the efficient hopping of the charge carriers~\cite{Hoofman1998,Sirringhaus1999}. 

In conclusion, we have shown that, when the temperature vanishes, the quantistic zero--point motion of the atoms
induces large corrections to the opto--electronic 
properties of Diamond and \tpa. The electronic bands are replaced by a continuum of polaronic states composed by 
entangled electron--phonon pairs. This successfully explains the 
sub--gap states observed experimentally 
in the absorption spectrum of Diamond, and the rich structures that appear in the \Sfs\,\,of \tpa.
The cooperative dynamics between electrons and atoms  that leads to the formation of the
polaronic states 
rules out any description in terms of
bare atoms, bare electronic states or quasi--particles.
This inevitably leads to the failure of the band--theory. 
The present results highlight the limitations of an approach to carbon--based nano--structures
that neglects the effect of the \epc. The deep change of the electronic and 
atomic wave--functions induced by the formation of the polaronic states can have important implications on related properties.
This inevitably, imposes a critical revision of the results obtained using purely electronic theories with
atoms frozen in their equilibrium positions.

The authors acknowledge C. Attaccalite for a critical reading of the manuscript.
This work was supported by the EU through the FP6 Nanoquanta NoE
(NMP4-CT-2004-50019), the FP7 ETSF I3 e-Infrastructure (Grant Agreement
211956). One of the authors (AM) would like to acknowledge support from 
the HPC-Europa2 Transnational collaboration project.

\bibliography{PRL}

\begin{thebibliography}{21}%
\makeatletter
\providecommand \@ifxundefined [1]{%
 \@ifx{#1\undefined}
}%
\providecommand \@ifnum [1]{%
 \ifnum #1\expandafter \@firstoftwo
 \else \expandafter \@secondoftwo
 \fi
}%
\providecommand \@ifx [1]{%
 \ifx #1\expandafter \@firstoftwo
 \else \expandafter \@secondoftwo
 \fi
}%
\providecommand \natexlab [1]{#1}%
\providecommand \enquote  [1]{``#1''}%
\providecommand \bibnamefont  [1]{#1}%
\providecommand \bibfnamefont [1]{#1}%
\providecommand \citenamefont [1]{#1}%
\providecommand \href@noop [0]{\@secondoftwo}%
\providecommand \href [0]{\begingroup \@sanitize@url \@href}%
\providecommand \@href[1]{\@@startlink{#1}\@@href}%
\providecommand \@@href[1]{\endgroup#1\@@endlink}%
\providecommand \@sanitize@url [0]{\catcode `\\12\catcode `\$12\catcode
  `\&12\catcode `\#12\catcode `\^12\catcode `\_12\catcode `\%12\relax}%
\providecommand \@@startlink[1]{}%
\providecommand \@@endlink[0]{}%
\providecommand \url  [0]{\begingroup\@sanitize@url \@url }%
\providecommand \@url [1]{\endgroup\@href {#1}{\urlprefix }}%
\providecommand \urlprefix  [0]{URL }%
\providecommand \Eprint [0]{\href }%
\@ifxundefined \urlstyle {%
  \providecommand \doi  [0]{\begingroup \@sanitize@url \@doi}%
  \providecommand \@doi [1]{\endgroup \@@startlink {\doibase
  #1}doi:\discretionary {}{}{}#1\@@endlink }%
}{%
  \providecommand \doi  [0]{doi:\discretionary{}{}{}\begingroup
  \urlstyle{rm}\Url }%
}%
\providecommand \doibase [0]{http://dx.doi.org/}%
\providecommand \Doi [0]{\begingroup \@sanitize@url \@Doi }%
\providecommand \@Doi  [1]{\endgroup\@@startlink{\doibase#1}\@@Doi}%
\providecommand \@@Doi [1]{#1\@@endlink}%
\providecommand \selectlanguage [0]{\@gobble}%
\providecommand \bibinfo  [0]{\@secondoftwo}%
\providecommand \bibfield  [0]{\@secondoftwo}%
\providecommand \translation [1]{[#1]}%
\providecommand \BibitemOpen [0]{}%
\providecommand \bibitemStop [0]{}%
\providecommand \bibitemNoStop [0]{.\EOS\space}%
\providecommand \EOS [0]{\spacefactor3000\relax}%
\providecommand \BibitemShut  [1]{\csname bibitem#1\endcsname}%
\bibitem [{\citenamefont {Kersting}\ \emph {et~al.}(1993)\citenamefont
  {Kersting}, \citenamefont {Lemmer}, \citenamefont {Mahrt}, \citenamefont
  {Leo}, \citenamefont {Kurz}, \citenamefont {B{\"a}ssler},\ and\ \citenamefont
  {G{\"o}bel}}]{Kersting1993}%
  \BibitemOpen
  \bibfield  {author} {\bibinfo {author} {\bibfnamefont {R.}~\bibnamefont
  {Kersting}}, \bibinfo {author} {\bibfnamefont {U.}~\bibnamefont {Lemmer}},
  \bibinfo {author} {\bibfnamefont {R.~F.}\ \bibnamefont {Mahrt}}, \bibinfo
  {author} {\bibfnamefont {K.}~\bibnamefont {Leo}}, \bibinfo {author}
  {\bibfnamefont {H.}~\bibnamefont {Kurz}}, \bibinfo {author} {\bibfnamefont
  {H.}~\bibnamefont {B{\"a}ssler}}, \ and\ \bibinfo {author} {\bibfnamefont
  {E.~O.}\ \bibnamefont {G{\"o}bel}},\ }\Doi {10.1103/PhysRevLett.70.3820}
  {\bibfield  {journal} {\bibinfo  {journal} {Phys. Rev. Lett.},\ }\textbf
  {\bibinfo {volume} {70}},\ \bibinfo {pages} {3820} (\bibinfo {year}
  {1993})}\BibitemShut {NoStop}%
\bibitem [{\citenamefont {Bradley}\ \emph {et~al.}(1990)\citenamefont
  {Bradley}, \citenamefont {Brown}, \citenamefont {Marks}, \citenamefont
  {Mackay}, \citenamefont {Friend}, \citenamefont {Burns},\ and\ \citenamefont
  {Holmes}}]{J.H.Burroughes1990}%
  \BibitemOpen
  \bibfield  {author} {\bibinfo {author} {\bibfnamefont {D.}~\bibnamefont
  {Bradley}}, \bibinfo {author} {\bibfnamefont {A.}~\bibnamefont {Brown}},
  \bibinfo {author} {\bibfnamefont {R.}~\bibnamefont {Marks}}, \bibinfo
  {author} {\bibfnamefont {K.}~\bibnamefont {Mackay}}, \bibinfo {author}
  {\bibfnamefont {R.}~\bibnamefont {Friend}}, \bibinfo {author} {\bibfnamefont
  {P.}~\bibnamefont {Burns}}, \ and\ \bibinfo {author} {\bibfnamefont
  {A.}~\bibnamefont {Holmes}},\ }\Doi {http://dx.doi.org/10.1038/347539a0}
  {\bibfield  {journal} {\bibinfo  {journal} {Nature},\ }\textbf {\bibinfo
  {volume} {347}},\ \bibinfo {pages} {539} (\bibinfo {year}
  {1990})}\BibitemShut {NoStop}%
\bibitem [{\citenamefont {Martini}\ \emph {et~al.}(2007)\citenamefont
  {Martini}, \citenamefont {Craig}, \citenamefont {Molenkamp}, \citenamefont
  {Miyata}, \citenamefont {Tolbert},\ and\ \citenamefont
  {Schwartz}}]{IgnacioB.Martini2007}%
  \BibitemOpen
  \bibfield  {author} {\bibinfo {author} {\bibfnamefont {I.~B.}\ \bibnamefont
  {Martini}}, \bibinfo {author} {\bibfnamefont {I.~M.}\ \bibnamefont {Craig}},
  \bibinfo {author} {\bibfnamefont {W.~C.}\ \bibnamefont {Molenkamp}}, \bibinfo
  {author} {\bibfnamefont {H.}~\bibnamefont {Miyata}}, \bibinfo {author}
  {\bibfnamefont {S.~H.}\ \bibnamefont {Tolbert}}, \ and\ \bibinfo {author}
  {\bibfnamefont {B.~J.}\ \bibnamefont {Schwartz}},\ }\Doi
  {http://dx.doi.org/10.1038/nnano.2007.294} {\bibfield  {journal} {\bibinfo
  {journal} {Nature Nanotechnology},\ }\textbf {\bibinfo {volume} {2}},\
  \bibinfo {pages} {647} (\bibinfo {year} {2007})}\BibitemShut {NoStop}%
\bibitem [{\citenamefont {Hannewald}\ \emph {et~al.}(2004)\citenamefont
  {Hannewald}, \citenamefont {{Stojanovi\ifmmode \acute{c}\else {\'c}\fi{}}},
  \citenamefont {Schellekens}, \citenamefont {Bobbert}, \citenamefont
  {Kresse},\ and\ \citenamefont {Hafner}}]{Hannewald2004}%
  \BibitemOpen
  \bibfield  {author} {\bibinfo {author} {\bibfnamefont {K.}~\bibnamefont
  {Hannewald}}, \bibinfo {author} {\bibfnamefont {V.~M.}\ \bibnamefont
  {{Stojanovi\ifmmode \acute{c}\else {\'c}\fi{}}}}, \bibinfo {author}
  {\bibfnamefont {J.~M.~T.}\ \bibnamefont {Schellekens}}, \bibinfo {author}
  {\bibfnamefont {P.~A.}\ \bibnamefont {Bobbert}}, \bibinfo {author}
  {\bibfnamefont {G.}~\bibnamefont {Kresse}}, \ and\ \bibinfo {author}
  {\bibfnamefont {J.}~\bibnamefont {Hafner}},\ }\Doi
  {10.1103/PhysRevB.69.075211} {\bibfield  {journal} {\bibinfo  {journal}
  {Phys. Rev. B},\ }\textbf {\bibinfo {volume} {69}},\ \bibinfo {pages}
  {075211} (\bibinfo {year} {2004})}\BibitemShut {NoStop}%
\bibitem [{\citenamefont {Onida}\ \emph {et~al.}(2002)\citenamefont {Onida},
  \citenamefont {Reining},\ and\ \citenamefont {Rubio}}]{Onida2002}%
  \BibitemOpen
  \bibfield  {author} {\bibinfo {author} {\bibfnamefont {G.}~\bibnamefont
  {Onida}}, \bibinfo {author} {\bibfnamefont {L.}~\bibnamefont {Reining}}, \
  and\ \bibinfo {author} {\bibfnamefont {A.}~\bibnamefont {Rubio}},\ }\Doi
  {10.1103/RevModPhys.74.601} {\bibfield  {journal} {\bibinfo  {journal} {Rev.
  Mod. Phys.},\ }\textbf {\bibinfo {volume} {74}},\ \bibinfo {pages} {601}
  (\bibinfo {year} {2002})}\BibitemShut {NoStop}%
\bibitem [{\citenamefont {Mahan}(1998)}]{mahan}%
  \BibitemOpen
  \bibfield  {author} {\bibinfo {author} {\bibfnamefont {G.}~\bibnamefont
  {Mahan}},\ }\href@noop {} {\emph {\bibinfo {title} {Many-Particle Physics}}}\
  (\bibinfo  {publisher} {(New York: Plenum)},\ \bibinfo {year}
  {1998})\BibitemShut {NoStop}%
\bibitem [{\citenamefont {Cardona}(2006)}]{Cardona2006}%
  \BibitemOpen
  \bibfield  {author} {\bibinfo {author} {\bibfnamefont {M.}~\bibnamefont
  {Cardona}},\ }\Doi {10.1016/j.stam.2006.03.009} {\bibfield  {journal}
  {\bibinfo  {journal} {Sci. Technol. Adv. Mater.},\ }\textbf {\bibinfo
  {volume} {7}},\ \bibinfo {pages} {S60} (\bibinfo {year} {2006})}\BibitemShut
  {NoStop}%
\bibitem [{\citenamefont {Marini}(2008)}]{Marini2008}%
  \BibitemOpen
  \bibfield  {author} {\bibinfo {author} {\bibfnamefont {A.}~\bibnamefont
  {Marini}},\ }\Doi {10.1103/PhysRevLett.101.106405} {\bibfield  {journal}
  {\bibinfo  {journal} {Phys. Rev. Lett.},\ }\textbf {\bibinfo {volume}
  {101}},\ \bibinfo {pages} {106405} (\bibinfo {year} {2008})}\BibitemShut
  {NoStop}%
\bibitem [{\citenamefont {Capaz}\ \emph {et~al.}(2005)\citenamefont {Capaz},
  \citenamefont {Spataru}, \citenamefont {Tangney}, \citenamefont {Cohen},\
  and\ \citenamefont {Louie}}]{Capaz2005}%
  \BibitemOpen
  \bibfield  {author} {\bibinfo {author} {\bibfnamefont {R.~B.}\ \bibnamefont
  {Capaz}}, \bibinfo {author} {\bibfnamefont {C.~D.}\ \bibnamefont {Spataru}},
  \bibinfo {author} {\bibfnamefont {P.}~\bibnamefont {Tangney}}, \bibinfo
  {author} {\bibfnamefont {M.~L.}\ \bibnamefont {Cohen}}, \ and\ \bibinfo
  {author} {\bibfnamefont {S.~G.}\ \bibnamefont {Louie}},\ }\Doi
  {10.1103/PhysRevLett.94.036801} {\bibfield  {journal} {\bibinfo  {journal}
  {Phys. Rev. Lett.},\ }\textbf {\bibinfo {volume} {94}},\ \bibinfo {pages}
  {036801} (\bibinfo {year} {2005})}\BibitemShut {NoStop}%
\bibitem [{\citenamefont {Giustino}\ \emph {et~al.}(2010)\citenamefont
  {Giustino}, \citenamefont {Louie},\ and\ \citenamefont
  {Cohen}}]{Giustino2010}%
  \BibitemOpen
  \bibfield  {author} {\bibinfo {author} {\bibfnamefont {F.}~\bibnamefont
  {Giustino}}, \bibinfo {author} {\bibfnamefont {S.~G.}\ \bibnamefont {Louie}},
  \ and\ \bibinfo {author} {\bibfnamefont {M.~L.}\ \bibnamefont {Cohen}},\
  }\Doi {10.1103/PhysRevLett.105.265501} {\bibfield  {journal} {\bibinfo
  {journal} {Phys. Rev. Lett.},\ }\textbf {\bibinfo {volume} {105}},\ \bibinfo
  {pages} {265501} (\bibinfo {year} {2010})}\BibitemShut {NoStop}%
\bibitem [{\citenamefont {Zollner}\ \emph {et~al.}(1992)\citenamefont
  {Zollner}, \citenamefont {Cardona},\ and\ \citenamefont
  {Gopalan}}]{Zollner1992}%
  \BibitemOpen
  \bibfield  {author} {\bibinfo {author} {\bibfnamefont {S.}~\bibnamefont
  {Zollner}}, \bibinfo {author} {\bibfnamefont {M.}~\bibnamefont {Cardona}}, \
  and\ \bibinfo {author} {\bibfnamefont {S.}~\bibnamefont {Gopalan}},\ }\Doi
  {10.1103/PhysRevB.45.3376} {\bibfield  {journal} {\bibinfo  {journal} {Phys.
  Rev. B},\ }\textbf {\bibinfo {volume} {45}},\ \bibinfo {pages} {3376}
  (\bibinfo {year} {1992})}\BibitemShut {NoStop}%
\bibitem [{\citenamefont {Baroni}\ \emph {et~al.}(2001)\citenamefont {Baroni},
  \citenamefont {de~Gironcoli}, \citenamefont {{Dal Corso}},\ and\
  \citenamefont {Giannozzi}}]{baroni2001}%
  \BibitemOpen
  \bibfield  {author} {\bibinfo {author} {\bibfnamefont {S.}~\bibnamefont
  {Baroni}}, \bibinfo {author} {\bibfnamefont {S.}~\bibnamefont
  {de~Gironcoli}}, \bibinfo {author} {\bibfnamefont {A.}~\bibnamefont {{Dal
  Corso}}}, \ and\ \bibinfo {author} {\bibfnamefont {P.}~\bibnamefont
  {Giannozzi}},\ }\Doi {10.1103/RevModPhys.73.515} {\bibfield  {journal}
  {\bibinfo  {journal} {Rev. Mod. Phys.},\ }\textbf {\bibinfo {volume} {73}},\
  \bibinfo {pages} {515} (\bibinfo {year} {2001})}\BibitemShut {NoStop}%
\bibitem [{\citenamefont {Dolcher}\ \emph {et~al.}(1992)\citenamefont
  {Dolcher}, \citenamefont {Grosso},\ and\ \citenamefont {Parravicini}}]{fano}%
  \BibitemOpen
  \bibfield  {author} {\bibinfo {author} {\bibfnamefont {V.}~\bibnamefont
  {Dolcher}}, \bibinfo {author} {\bibfnamefont {G.}~\bibnamefont {Grosso}}, \
  and\ \bibinfo {author} {\bibfnamefont {G.~P.}\ \bibnamefont {Parravicini}},\
  }\Doi {10.1103/PhysRevB.46.9312} {\bibfield  {journal} {\bibinfo  {journal}
  {Phys. Rev. B},\ }\textbf {\bibinfo {volume} {46}},\ \bibinfo {pages} {9312}
  (\bibinfo {year} {1992})}\BibitemShut {NoStop}%
\bibitem [{\citenamefont {Cannuccia}\ and\ \citenamefont
  {Marini}()}]{cannuccia_u}%
  \BibitemOpen
  \bibfield  {author} {\bibinfo {author} {\bibfnamefont {E.}~\bibnamefont
  {Cannuccia}}\ and\ \bibinfo {author} {\bibfnamefont {A.}~\bibnamefont
  {Marini}},\ }\href@noop {} {\bibfield  {journal} {\bibinfo  {journal}
  {(unpublished)}},\ }\bibinfo {note} {it can be demonstrated that the $\left|
  P\right. \ra$ states are eigenstates of the fictitious Hamiltonian
  $H_{\(I\gl\)\(J\gl'\)}=\gd_{IJ}(\gee_I+\gS^{DW}_I)+\gd_{\gl\gl'}\go_{\gl}+g_%
{IJ}^{\gql}$}\BibitemShut {NoStop}%
\bibitem [{\citenamefont {Logothetidis}\ \emph {et~al.}(1992)\citenamefont
  {Logothetidis}, \citenamefont {Petalas}, \citenamefont {Polatoglou},\ and\
  \citenamefont {Fuchs}}]{Logothetidis1992}%
  \BibitemOpen
  \bibfield  {author} {\bibinfo {author} {\bibfnamefont {S.}~\bibnamefont
  {Logothetidis}}, \bibinfo {author} {\bibfnamefont {J.}~\bibnamefont
  {Petalas}}, \bibinfo {author} {\bibfnamefont {H.~M.}\ \bibnamefont
  {Polatoglou}}, \ and\ \bibinfo {author} {\bibfnamefont {D.}~\bibnamefont
  {Fuchs}},\ }\Doi {10.1103/PhysRevB.46.4483} {\bibfield  {journal} {\bibinfo
  {journal} {Phys. Rev. B},\ }\textbf {\bibinfo {volume} {46}},\ \bibinfo
  {pages} {4483} (\bibinfo {year} {1992})}\BibitemShut {NoStop}%
\bibitem [{cal()}]{calculations}%
  \BibitemOpen
  \href@noop {} {}\bibinfo {note} {The phonon modes and the electron--phonon
  matrix elements were calculated using a uniform grid of $4\times 4\times 4$
  and $10 \times 1 \times 1$ k--points for Diamond and {\it
  trans}--polyacetylene respectively. We used a plane--waves basis and N.
  Troullier and J. L. Martins pseudopotentials for the carbon and hydrogen
  atoms. For the ground--state calculations we used the \texttt{PWSCF}
  code\,\cite{pwscf}. The Fan self-energy and the Debye--Waller contribution
  are calculated using a random grid of transferred momenta, using the
  \texttt{yambo} code\,\cite{Marini20091392}}\BibitemShut {NoStop}%
\bibitem [{\citenamefont {Heeger}\ \emph {et~al.}(1988)\citenamefont {Heeger},
  \citenamefont {Kivelson}, \citenamefont {Schrieffer},\ and\ \citenamefont
  {Su}}]{HeegerReview}%
  \BibitemOpen
  \bibfield  {author} {\bibinfo {author} {\bibfnamefont {A.~J.}\ \bibnamefont
  {Heeger}}, \bibinfo {author} {\bibfnamefont {S.}~\bibnamefont {Kivelson}},
  \bibinfo {author} {\bibfnamefont {J.~R.}\ \bibnamefont {Schrieffer}}, \ and\
  \bibinfo {author} {\bibfnamefont {W.~P.}\ \bibnamefont {Su}},\ }\Doi
  {10.1103/RevModPhys.60.781} {\bibfield  {journal} {\bibinfo  {journal} {Rev.
  Mod. Phys.},\ }\textbf {\bibinfo {volume} {60}},\ \bibinfo {pages} {781}
  (\bibinfo {year} {1988})}\BibitemShut {NoStop}%
\bibitem [{\citenamefont {Hoofman}(1998)}]{Hoofman1998}%
  \BibitemOpen
  \bibfield  {author} {\bibinfo {author} {\bibfnamefont {R.~J.}\ \bibnamefont
  {Hoofman}},\ }\Doi {http://dx.doi.org/10.1038/32118} {\bibfield  {journal}
  {\bibinfo  {journal} {Nature},\ }\textbf {\bibinfo {volume} {392}},\ \bibinfo
  {pages} {54} (\bibinfo {year} {1998})}\BibitemShut {NoStop}%
\bibitem [{\citenamefont {Sirringhaus}\ \emph {et~al.}(1999)\citenamefont
  {Sirringhaus}, \citenamefont {Brown}, \citenamefont {Friend}, \citenamefont
  {Nielsen}, \citenamefont {Bechgaard}, \citenamefont {Langeveld-Voss},
  \citenamefont {Spiering}, \citenamefont {Janssen},\ and\ \citenamefont
  {de~Leeuw~Meijer}}]{Sirringhaus1999}%
  \BibitemOpen
  \bibfield  {author} {\bibinfo {author} {\bibfnamefont {P.~J.}\ \bibnamefont
  {Sirringhaus}}, \bibinfo {author} {\bibfnamefont {R.~H.}\ \bibnamefont
  {Brown}}, \bibinfo {author} {\bibfnamefont {M.~M.}\ \bibnamefont {Friend}},
  \bibinfo {author} {\bibfnamefont {K.}~\bibnamefont {Nielsen}}, \bibinfo
  {author} {\bibfnamefont {B.~M.~W.}\ \bibnamefont {Bechgaard}}, \bibinfo
  {author} {\bibfnamefont {A.~J.~H.}\ \bibnamefont {Langeveld-Voss}}, \bibinfo
  {author} {\bibfnamefont {R.~A.~J.}\ \bibnamefont {Spiering}}, \bibinfo
  {author} {\bibfnamefont {E.~W.}\ \bibnamefont {Janssen}}, \ and\ \bibinfo
  {author} {\bibfnamefont {P.~H. D.~M.}\ \bibnamefont {de~Leeuw~Meijer}},\
  }\Doi {http://dx.doi.org/10.1038/44359} {\bibfield  {journal} {\bibinfo
  {journal} {Nature},\ }\textbf {\bibinfo {volume} {401}},\ \bibinfo {pages}
  {685} (\bibinfo {year} {1999})}\BibitemShut {NoStop}%
\bibitem [{\citenamefont {Giannozzi}\ and\ \citenamefont {al.}(2009)}]{pwscf}%
  \BibitemOpen
  \bibfield  {author} {\bibinfo {author} {\bibfnamefont {P.}~\bibnamefont
  {Giannozzi}}\ and\ \bibinfo {author} {\bibnamefont {al.}},\ }\href@noop {}
  {\bibfield  {journal} {\bibinfo  {journal} {J. Phys. Condens. Matter},\
  }\textbf {\bibinfo {volume} {21}},\ \bibinfo {pages} {395502} (\bibinfo
  {year} {2009})}\BibitemShut {NoStop}%
\bibitem [{\citenamefont {Marini}\ \emph {et~al.}(2009)\citenamefont {Marini},
  \citenamefont {Hogan}, \citenamefont {Gr{\"u}ning},\ and\ \citenamefont
  {Varsano}}]{Marini20091392}%
  \BibitemOpen
  \bibfield  {author} {\bibinfo {author} {\bibfnamefont {A.}~\bibnamefont
  {Marini}}, \bibinfo {author} {\bibfnamefont {C.}~\bibnamefont {Hogan}},
  \bibinfo {author} {\bibfnamefont {M.}~\bibnamefont {Gr{\"u}ning}}, \ and\
  \bibinfo {author} {\bibfnamefont {D.}~\bibnamefont {Varsano}},\ }\href@noop
  {} {\bibfield  {journal} {\bibinfo  {journal} {Computer Physics
  Communications},\ }\textbf {\bibinfo {volume} {180}},\ \bibinfo {pages}
  {1392} (\bibinfo {year} {2009})}\BibitemShut {NoStop}%
\end{thebibliography}%

\end{document}